# Contrast in transmission spectroscopy of a single quantum dot

B. D. Gerardot[*], S. Seidl[*], P. A. Dalgarno, and R. J. Warburton
*School of Engineering and Physical Sciences, Heriot-Watt University, Edinburgh EH14 4AS, UK*

M. Kroner and K. Karrai
*Center for NanoScience and Department für Physik der LMU, Geschwister-Scholl-Platz 1, 80539 Munich, Germany*

A. Badolato and P. M. Petroff,
*Materials Department, University of California, Santa Barbara, California 93106*

We perform transmission spectroscopy on single quantum dots and examine the effects of a resident carrier's spin, the incident laser spot size, polarization, and power on the experimental contrast. We demonstrate a factor of 4 improvement in the maximum contrast by using a solid immersion lens to decrease the spot area. This increase yields a maximum signal to noise ratio of ~ 2000 Hz$^{-1/2}$, which will allow for MHz detection frequencies. We anticipate that this improvement will allow further investigation of spectral fluctuation and open up the feasibility for an all-optical read-out of an electron spin in a quantum dot.

The ability to process quantum information requires access to a highly coherent two-level system [1]. Zero-dimensional semiconductor quantum dots (QDs) are an attractive solid-state host for such systems due to their reduced spin relaxation [2] and decoherence mechanisms compared to bulk or quantum wells. An optically driven exciton [3] or the spin state of a single carrier [1] in a QD are two such candidates. Photoluminescence (PL) spectroscopy is commonly used to characterize optically active QDs. In this method, carriers generated above the ground state of the QD relax to the lowest energy excited state through a generally incoherent process. While this technique is useful for probing the energy levels, charged states, and fine-structure of a QD, the relaxation process and the presence of other carriers in excited states introduces decoherence.

A complimentary experimental approach to PL is laser spectroscopy, in which the differential transmission of a single QD as a function of laser detuning is measured [4-7]. Laser spectroscopy provides two significant advantages over PL: sub-µeV resolution and the well-defined preparation of states in a QD. The high-resolution provides access to the true lineshape and linewidth of the transition. Investigation of the lineshape can indicate interference effects [8,9] whereas deviations from the homogeneous linewidth can be caused by a non-radiative processes such as tunnelling [7] or spectral fluctuations due to environmental effects [6]. The well-defined preparation of states in transmission spectroscopy is a result of the good optical selection rules in the QD owing to the inherent lattice symmetry. Resonant excitation of a transition will transfer the polarization of the incident laser light to the exciton, allowing preparation of the individual electron and hole spins. This capability permits the optical initialization of single electron spins [10] and offers promise for readout of the electron spin [11].

The maximum change in transmission (i.e. the contrast) is a measurement of the ratio of light resonantly scattered by the dot compared to the incident laser intensity. However, detecting the resonantly scattered laser light from a self-assembled QD is a non-trivial experiment. Here we explore how the polarization, power, and spot size of the incident laser all affect the contrast. By using a solid immersion lens (SIL), we achieve a spatial resolution of ~ 350 nm (FWHM) using 950 nm wavelength light. This increased resolution yields an increase in the experimental contrast by a factor of 4.8, and a maximum contrast of ~ 6% is observed in our system. We also demonstrate that the spin state of a resident electron in the QD contributes to the ultimate contrast. Current measurements use integration times of 0.1 – 1.0 second, but motivated by the exciting possibility to directly observe spectral fluctuations or carrier spin-flips, we consider the suitability for µ-second measurement times in transmission spectroscopy of a single QD.

We use a confocal microscope with diffraction-limited resolution to perform both PL and transmission spectroscopy. Our system's spatial resolution was determined by focussing a laser beam on a 10 µm period Al/glass grating and measuring the differential transmission as it moves across the Al/glass interface [Fig. 1(a)]. The distance travelled was calibrated with a Fabry-Perot cavity using a laser of known wavelength. The FWHM of the spot diameter ($\Delta x$) without (with) a SIL was measured to be 755 (344) nm at a wavelength ($\lambda$) of 950 nm. For a focused plane wave, $\Delta x_{FWHM} = (0.52*\lambda)/(NA_{obj}*n)$, where $NA_{obj}$ is the numerical aperture of the objective lens and $n$ is the refractive index of the material. For $\lambda = 950$ nm, $NA_{obj} = 0.65$, and $n = 1.0$ (air), $\Delta x_{FWHM}$ is 760 nm. Using a glass ($n = 2.0$ for $\lambda = 950$ nm) hemispherical SIL directly on top of the semiconductor sample surface, $\Delta x_{FWHM} = 390$ nm. This criterion is valid for plane waves; the use of Gaussian optics in the setup could account for the difference between predicted and measured values, but the visibility of Airy rings in the data support the plane wave approximation. Nevertheless, the experiment demonstrates diffraction limited performance for our system we measure that using a glass SIL reduces the spot area by 4.8x.

Our sample grown by molecular beam epitaxy consists of InGaAs QDs embedded in a MISFET device [6]. The QDs are located 136 nm below the GaAs-vacuum interface to ensure the resonance is absorptive rather than dispersive [9]. We select the QD charge state due to Coulomb blockade by applying an electrical bias between a top Schottky gate and a back Ohmic contact. The QD energy is modulated via the Stark shift and a lock-in amplifier is used to measure the differential transmission [12]. A Si p-i-n detector (Thorlabs FDS 100) is mounted directly below the sample at 4K inside the cryostat and a current amplifier (Femto DLPCA-200) is positioned near the cryostat to amplify the signal sent to the lock-in amplifier.

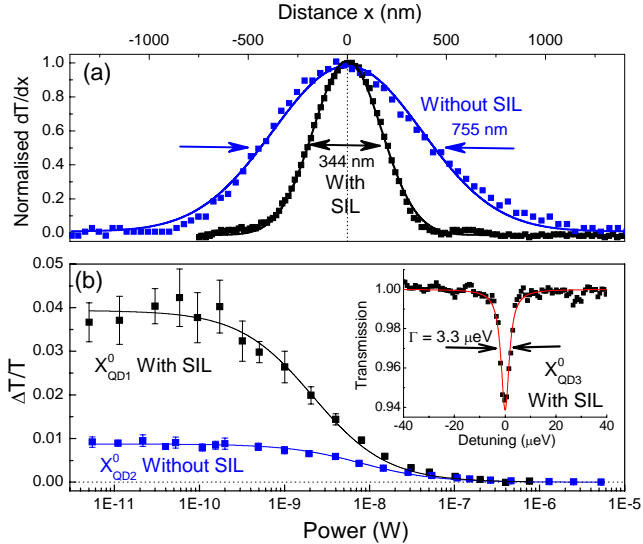

FIG. 1: (a) The spatial resolution is determined by measuring the differential transmission of the focused laser spot as it traverses a metal/glass interface (solid lines are Gaussian fits to the data). (b) Saturation curves for two different $X^0$ transitions with and without the use of a SIL. In the low power regime a SIL improves the contrast by 4.1±.01x. At higher powers the curves converge due to saturation as expected for a two-level system (solid lines). The inset shows the transmission of a different $X^0$ transition as a function of detuning. Here, the contrast is 6.02% and the measured linewidth is 3.3 µeV due to a spectral fluctuation.

Fig. 1(b) shows typical results for the contrast of a neutral exciton ($X^0$) on and off the SIL in two different QDs. The measurement, over 6 decades in power, is made using linearly polarized light to maximize the contrast (see Fig. 2 inset). Due to the electron-hole exchange interaction, the bright exciton lines are non-degenerate and linearly polarized [13]. We have characterized the contrast of ~ 30 different QDs. The statistics on all QDs measured show a contrast in the low power regime of 3.7±0.7% (0.77±0.3%) with(out) a SIL, demonstrating that the contrast is improved by a factor of 4.8±0.01 due to the reduction in spot size. The vacuum level / first excited state transition of a QD behaves like a two-level system due to its atomic-like character. Therefore, absorption of the transition will saturate in the limit of a strong driving field. Saturation results in a decreased contrast at resonance and an increase in the linewidth, or power broadening. Below the saturation power the maximum contrast is determined by the scattering cross-section divided by the incident laser spot area. It is notable that the measured values are still ~ 6x less than predicted from the scattering cross-section and spot area [9]. Naturally, further improvements are possible using a SIL with a larger $n$ or of the Weierstrass geometry [14].

The inset of Fig. 1(b) shows the transmission of an $X^0$ in a different QD with 6% contrast, the largest contrast we have observed to date. The measured linewidth, $\Gamma$, for this transition is 3.3 µeV. Using time-resolved PL, we have directly measured a radiative recombination lifetime of 0.75 nsec for this $X^0$ (QD3), corresponding to a homogeneous linewidth of 0.88 µeV. In this sample, we measure in transmission a distribution for $\Gamma$ between 0.9 and 4.5 µeV. The additional broadening mechanism has been attributed to temporal fluctuations in the resonance position of the QD [6]. The origin of such a mechanism has not yet been verified; possibilities include electrostatic fluctuations in the back contact or nearby QDs. Measurements with integration times (bandwidths) between 0.05 (3.3) and 50 second (0.003 Hz) yield the same $\Gamma$, suggesting that the fluctuation is on a sub-ms timescale. Thus, access to a smaller integration time is needed to characterize the spectral diffusion and identify ways to eliminate it.

Figure 2 compares the saturation curves for $X^0$ and $X^{1-}$ from the same QD. The oscillator strengths are nearly equal [15], yet the maximum contrast of $X^0$ in the limit of low power is twice that of $X^{1-}$. In addition to the incident laser spot size and polarization, the spin state of any resident carriers in the QD affects the contrast. The ground electron (hole) states in a QD are degenerate with respect to their spin projection ± 1/2 (± 3/2) at zero magnetic field. Resonant excitation of a singly negatively charged QD creates a trion consisting of two electrons and a hole ($X^{1-}$), for which there is zero total electron spin and hence the electron-hole exchange interaction is turned off. However, due to the Pauli principle, the polarization of the optical excitation must excite an electron opposite in spin to the resident electron. In our devices, when we apply the appropriate bias, an unpolarized electron tunnels into the QD from the back gate. Also, at zero magnetic field, the electron spin relaxation time is much shorter than the integration time used in the measurement, ensuring that the electron has a random spin orientation [16]. Hence, there is a 50% probability that the optical excitation will create the $X^{1-}$. Fig. 2 confirms that transmission spectroscopy is sensitive to the spin of a resident electron in a QD, but the relatively long integration time measures an average of the carrier spin. Statistically, this is equivalent to an ensemble averaging. If the integration time can be made smaller than the electron spin relaxation time, the contrast would switch between two discrete levels in real time due to electron spin quantum jumps.

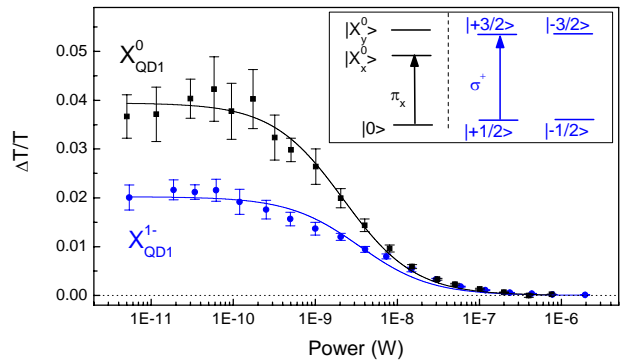

FIG. 2: Saturation curves for the $X^{1-}$ and $X^0$ transitions for the same QD. The inset shows level diagrams and the polarizations used in the measurements for the two transitions. The $X^0$ contrast is 1.85±.0x larger than the $X^{1-}$ contrast.

Spectral fluctuation and carrier spin flips in the QD occur on a much faster timescale than the required integration time, therefore a significant challenge in QD spectroscopy is to increase the detection frequency to resolve these individ-ual events. The integration time in our experiment is determined by the signal to noise ratio (*S/N*) of the system, where the *S* is the differential transmission at zero detuning and *N* the standard deviation of the measured transmission. The main experimental noise sources are: the detector, the current pre-amplifier, and the shot noise of the laser light. A useful figure of merit to quantify the device noise is the noise equivalent power (NEP), which specifies the root mean square (rms) value of a sinusoidally modulated input signal that yields an rms output signal equal to the rms noise in the device. The combined room temperature NEP of the detector and current pre-amplifier listed by the manufacturers is 13.9, 24.7, and 72.7 fW/Hz$^{0.5}$ at $10^9$, $10^8$, and $10^7$ amplification, respectively. The shot noise from the laser is defined as the square root of the incoming photon rate. Fig. 3(a) shows the NEP as a function of power measured off-resonance from any QD transition. The measurements were performed under nominally identical conditions; however they were taken over several days and the exact environment (i.e. electrical noise, mechanical vibrations, etc.) may have changed, which accounts for the scatter in the data. The minimum NEP at a certain excitation power can be considered the ideal system performance. In the low power regime the measured NEP is 4.5 fW/Hz$^{0.5}$, which likely is less than the expected room temperature NEP (13.9 fW/Hz$^{0.5}$) due to a reduction in the thermal noise of the photodiode. In the high power regime the NEP increases due to the decreased current pre-amplifier sensitivity and the photon noise.

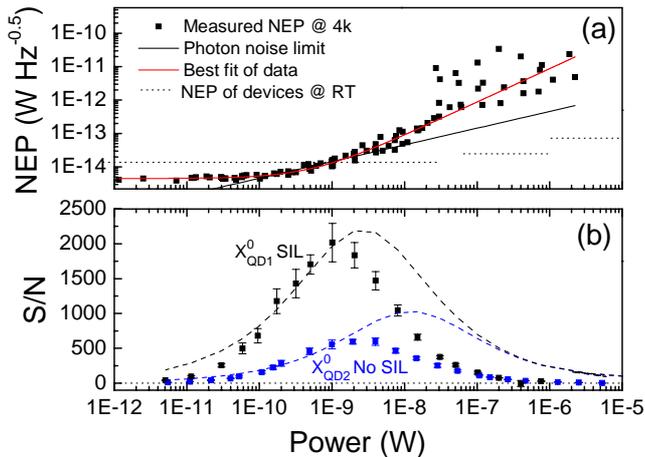

FIG. 3: (a) The NEP of our measurement system as a function of incident laser power. The data points were measured during different experimental runs with nominally the same setup. The straight solid line corresponds to the shot noise from the laser, the curved line is a best fit to the data, and the dashed lines are the combined room temperature NEP of the pre-amplifier and the detector. (b) *S/N* as function of incident laser power. The signal from Fig. 1b and the noise from the fit in (a) are used to determine the experimental *S/N*. The dashed lines correspond to the ideal case when the laser shot noise dominates.

Fig. 3(b) shows the *S/N* for the two QDs presented in Fig. 1(b), with and without a SIL. The noise used to determine the *S/N* in the plot is the best fit to the data points in Fig. 3(a) (*NEP*=$\alpha$+$\beta P$, where $\alpha$=4.5e-15 fW/Hz$^{0.5}$ and $\beta$=8.0e-6 Hz$^{-0.5}$), which represents the typical system noise. The dashed lines correspond to the idealized *S/N* with no noise from the equipment: *S* is calculated using the two-level model with the contrast in the low power regime from Fig. 1(b) and *N* is the laser shot noise. The maximum *S/N* when using the SIL (~2000/Hz$^{0.5}$) is approximately 4 times larger than without a SIL. The maximum occurs just before saturation begins for each curve. In the saturation regime, the *S/N* does not depend on the spot area. In the low power regime, the *S/N* decreases due to the decreasing signal but constant noise.

With a repetitive signal, averaging over many cycles improves the *S/N* in proportion to $\eta^{0.5}$, where $\eta$ is the number of cycles. Based on a *S/N* of 2000 at 1 Hz, one can expect a signal to still be observable (i.e. *S/N* = 1) at a 4 MHz bandwidth (~0.5 μs integration time). However, for the lock-in technique the modulation frequency generally should be 10x larger than the measurement frequency. The RC time constant of our device (~500 ns) is sufficient for MHz frequency modulation, but the bandwidth of our pre-amplifier is limited to 1.2 kHz using $10^9$ amplification. One strategy to circumvent the pre-amplifier bandwidth limit is perhaps to measure reflection rather than transmission [8,9]. In this case the detector can be located outside the cryostat at room temperature and an avalanche photodiode with internal gain can be used, thus reducing the required signal amplification.

In summary, we have examined how the size of the laser spot, the laser polarization, the incident laser power, and the spin state of a resident electron affect the contrast in transmission spectroscopy of a self-assembled QD. By using a SIL, we have demonstrated an increase in the maximum contrast and *S/N* by a factor of 4 due to the reduced spot size. The increase in the *S/N* makes measurements in the MHz frequency regime possible. This capability will allow further investigation into spectral fluctuations, hopefully leading to a strategy to eliminate them. Furthermore, real time measurement of electron spin flips in semiconductor QDs may become possible, allowing direct probing of spin relaxation dynamics.

We acknowledge financial support for this work from EPSRC (UK); SANDiE (EU); and DAAD and SFB631 (DE). *BDG and SS contributed equally to this work.